%% file: main.tex
\documentclass[aps,prl,reprint,superscriptaddress,nofootinbib]{revtex4-2}

\usepackage[T1]{fontenc}
\usepackage[utf8]{inputenc}
\usepackage{amsmath,amssymb,bm}
\usepackage{graphicx}
\usepackage{microtype}
\usepackage{tikz}
\usepackage[hidelinks]{hyperref}
\usetikzlibrary{arrows.meta,calc}

\newcommand{\dd}{\mathrm{d}}
\newcommand{\ee}{\mathrm{e}}
\newcommand{\ii}{\mathrm{i}}
\newcommand{\Tr}{\operatorname{Tr}}
\newcommand{\one}{\mathbf{1}}
\newcommand{\up}{\uparrow}
\newcommand{\dn}{\downarrow}
\newcommand{\Renyi}{R\'enyi}

\begin{document}

\title{Diagrammatic Monte Carlo for Fermionic \Renyi{} Entanglement Entropy}

\author{Boyuan Shi}
\affiliation{Department of Physics, Imperial College London,
London SW7 2AZ, United Kingdom}

\date{October 12, 2026}

\begin{abstract}
We develop a direct diagrammatic Monte Carlo method for the second
\Renyi{} entropy of interacting lattice fermions.  Fermionic replica
sewing yields a connected-determinant (CDet) expansion, which we sample
with many-configuration Markov-chain Monte Carlo.  On a $3\times3$
Hubbard cluster, deterministic coefficients through third order agree
with exact-diagonalization derivatives for three reference shifts,
while sampling extends the series to tenth order.  For a reference
shift with a continuous stability region, conformal-Borel resummation
also gives good agreement with the exact entropy.  We further compute
coefficients through ninth order on $40\times40$ and $50\times50$
periodic lattices, with subsystems of up to 676 sites.  The resummed
results from different reference shifts show good agreement at a
common truncation order.  Their shift and order dependence provide
checks on the stability of the continuation.
These calculations extend connected-diagram sampling to spatial
\Renyi{} observables on large finite fermionic systems.
\end{abstract}

\maketitle

\section{Introduction}

The scaling of entanglement entropy distinguishes gapped many-body states
from states with a Fermi surface.  Ground states of local gapped
Hamiltonians typically obey an area law; a Fermi surface adds a
logarithmic factor whose coefficient depends on its geometry
\cite{Eisert2010,Wolf2006,GioevKlich2006,Swingle2010}.  Following this
scaling through an interacting metallic or pseudogap regime requires
entropy calculations for several subsystem sizes.  For fermions, this
is difficult even when thermodynamic observables are accessible: a
replica estimator changes the imaginary-time boundary conditions, while
the underlying simulation may already have a severe sign problem
\cite{TroyerWiese2005,WangTroyer2014,BroeckerTrebst2014}.

Quantum Monte Carlo (QMC) replica and swap estimators have been developed
for spin, bosonic, and fermionic systems
\cite{Hastings2010,Humeniuk2012,Grover2013}.  Improvements in entropy
estimators have extended the accessible sizes and interaction regimes
\cite{Zhao2022PRL,Zhao2022NPJ,Pan2023,DEmidio2024,Jiang2025,WangYan2025}.
These methods are particularly effective when the sampling measure is
nonnegative; away from such regimes, the sign problem continues to
restrict finite-temperature calculations on large two-dimensional
fermionic lattices.

Diagrammatic Monte Carlo (DiagMC) samples terms of a Feynman expansion.
Connected-diagram methods have obtained results in correlated
Fermi-liquid, pairing, and pseudogap regimes
\cite{Kozik2010,Deng2015,Wu2017}.  Connected-determinant (CDet) summation
evaluates all connected diagrams at fixed interaction vertices through
subset recursions, combining cancellations before sampling
\cite{Rossi2017}.  CDet and related formulations have since reached high
orders in pseudogap and magnetic-transition calculations
\cite{Simkovic2024,Lenihan2022,Garioud2024}.  Many-configuration samplers
use the subset values to measure several orders in the same step
\cite{SimkovicRossi2021,Shi2025}.  Most applications concern
thermodynamic quantities or correlation functions.
Ref.~\cite{MoitraSensarma2023} expressed fermionic entropy in terms of
connected correlators, whereas we expand the replicated partition
function directly, with the spatial cut built into its free propagators.

Cold-atom experiments provide a further motivation.  Two-copy
interference has measured purity, \Renyi{} entropy, and mutual
information in an itinerant Bose--Hubbard system; microscope and
randomized-measurement protocols offer routes to fermionic Hubbard
systems \cite{Pichler2013,Islam2015,Elben2018}.  Fermi--Hubbard
simulators prepare hundreds of atoms, and numerical calculations help
determine their entropy and temperature \cite{Cocchi2017,Xu2025}.
Exact diagonalization (ED) supplies small-system benchmarks, while
determinantal QMC (DQMC) reaches large systems in favorable sign
regimes.  We seek a complementary calculation for a specified finite
lattice and subsystem in regimes with a severe sign problem.  We test
the replica expansion against ED on a small cluster, then examine its
shift and order dependence on lattices up to $50\times50$.

\section{Graded-replica expansion}

We consider the repulsive Hubbard model on an $N_1\times N_2$ periodic square
lattice, with nearest-neighbor hopping set to unity,
\begin{equation}
H=-\sum_{\langle ij\rangle\sigma}c_{i\sigma}^{\dagger}c_{j\sigma}
 +U\sum_i n_{i\up}n_{i\dn}-\mu\sum_{i\sigma}n_{i\sigma}.
\label{eq:hubbard}
\end{equation}
Here $c_{i\sigma}$ annihilates a fermion and
$n_{i\sigma}=c_{i\sigma}^\dagger c_{i\sigma}$; the hopping sum includes
both orientations of each nearest-neighbor bond.  For
$\rho=\ee^{-\beta H}/Z_1$, let $\bar A$ denote the complement of $A$,
$\rho_A=\Tr_{\bar A}\rho$, and $\Tr_A$ the trace over $A$.  Then
$Z_n[A]=\Tr_A\rho_A^n$ and $S_n[A]=(1-n)^{-1}\ln Z_n[A]$.  At finite
temperature this subsystem entropy contains thermal and quantum
contributions.  It becomes a bipartite entanglement entropy in the
nondegenerate ground-state limit $\beta\to\infty$.  Mutual-information
combinations can remove the leading bulk thermal term; negativity
addresses mixed-state entanglement \cite{WolfMutual2008,LuHsiehGrover2020}.
We calculate the full subsystem moment and report its intensive logarithm
\begin{equation}
f_A(\xi)=\frac{1}{N}\ln Z_2[A;\xi]
=f_A^{(0)}+\sum_{k\geq1}\frac{c_k}{N}\xi^k,
\qquad N=N_1N_2,
\label{eq:series}
\end{equation}
where $\xi$ interpolates a shifted homotopy Hamiltonian to the physical point
$\xi=1$:
\begin{align}
H(\xi)&=H_0'+\xi H_{\rm int}',\nonumber\\
H_0'&=H_{\rm kin}-\sum_{i\sigma}(\mu-\alpha_\sigma)n_{i\sigma},\nonumber\\
H_{\rm int}'&=U\sum_i n_{i\up}n_{i\dn}
-\sum_{i\sigma}\alpha_\sigma n_{i\sigma}.
\label{eq:homotopy}
\end{align}
Since $H(1)=H$, varying $\alpha_\sigma$ reorganizes the expansion around
different free references with the same physical endpoint.

The unnormalized replica moment is a graded twisted trace,
\begin{equation}
\widetilde Z_n[A]=\Tr_{\mathcal H^{\otimes n}}
\!\left[S_A^{\rm gr}(\ee^{-\beta H})^{\otimes n}\right],
\qquad Z_n[A]=\widetilde Z_n[A]/Z_1^n.
\label{eq:twisted}
\end{equation}
Here $S_A^{\rm gr}$ denotes the trace-compatible fermionic sewing
operator, including the parity compensation for even $n$.  Its
coherent-state representation fixes the temporal gluing.  With Grassmann
fields $\bar c,c$, the unshifted Hamiltonian has one-body matrix $h$ and
path integral
\begin{align}
\widetilde Z_n[A]
&=\int_{\mathrm{gr}\,A}\mathcal D[\bar c,c]\,
\ee^{-S_n[\bar c,c]},\nonumber\\
S_n&=\sum_{\alpha=1}^n\int_0^\beta\!\dd\tau\,
\left[\bar c^{(\alpha)}(\partial_\tau+h)c^{(\alpha)}
+U\sum_i n_{i\up}^{(\alpha)}n_{i\dn}^{(\alpha)}\right].
\label{eq:path}
\end{align}
On the complement $B=\bar A$, every sheet is antiperiodic.  On $A$, adjacent
sheets are cyclically joined and the last junction carries the fermionic
minus sign,
\begin{align}
c_B^{(\alpha)}(\beta^-)&=-c_B^{(\alpha)}(0^+),\nonumber\\
c_A^{(\alpha)}(\beta^-)&=c_A^{(\alpha+1)}(0^+),\quad \alpha<n,
&c_A^{(n)}(\beta^-)&=-c_A^{(1)}(0^+).
\label{eq:bc}
\end{align}
Figure~\ref{fig:gluing} summarizes the two boundary conditions for $n=2$.

\begin{figure}[t]
\centering
\input{fig_graded_swap_boundary_tikz.tex}
\caption{Two-sheeted graded replica geometry.  The colored slit is the
entangling region $A$ at the temporal cut.  Its banks are sewn cyclically
between replicas, with a minus sign on the final return; the complement
$\bar A$ closes antiperiodically within each sheet.}
\label{fig:gluing}
\end{figure}
Writing the boundary matrix as
$-P_B\otimes\one_n+P_A\otimes K_n$, one has $K_n^n=-\one_n$.  Hence the
replica momenta and sector twist are
\begin{equation}
q_m=\frac{(2m+1)\pi}{n},\qquad
\Omega_m=P_B-\ee^{\ii q_m}P_A.
\label{eq:q}
\end{equation}
For the convention $\bar G=\langle T_\tau c(\tau)c^\dagger(\tau')\rangle$,
define $A_m=[\one+\Omega_m^{-1}\ee^{-\beta h}]^{-1}$.  The exact free
sector propagator is
\begin{equation}
\bar G^{(m)}(\tau,\tau')=
\begin{cases}
\ee^{-\tau h}A_m\ee^{\tau'h},&\tau>\tau',\\
\ee^{-\tau h}(A_m-\one)\ee^{\tau'h},&\tau<\tau',
\end{cases}
\label{eq:green}
\end{equation}
and a discrete Fourier transform reconstructs the replica indices.
We tabulate each smooth time branch using a two-dimensional Chebyshev
expansion for repeated evaluation during sampling.

Let $X_\ell=(i_\ell,\tau_\ell,\alpha_\ell)$.  For the unshifted quartic
interaction, Wick's theorem gives, at order $k$,
\begin{align}
\frac{\widetilde Z_n^{(k)}}{\widetilde Z_{0,n}}
&=\frac{(-U)^k}{k!}\sum_{\{i_\ell,\alpha_\ell\}}
\int_0^\beta\prod_{\ell=1}^k\dd\tau_\ell\nonumber\\
&\quad\times\det\bar G_\up^{(n)}(X_p,X_q)
\det\bar G_\dn^{(n)}(X_p,X_q).
\label{eq:det}
\end{align}
In the shifted expansion, $h$ is replaced by the one-body matrix of
$H_0'$.  The up- and down-spin determinant diagonals receive
$\alpha_\dn/U$ and $\alpha_\up/U$, respectively, with the lower equal-time
contraction $\bar G(\tau,\tau^+)$.  The constant from this factorization
cancels in the normalized logarithm.
CDet subset recursions isolate $\ln\widetilde Z_n[A]$, after which
$\ln Z_n[A]=\ln\widetilde Z_n[A]-n\ln Z_1$.  We sample orders
simultaneously using many-configuration Markov-chain Monte Carlo (MCMCMC) of
Ref.~\cite{SimkovicRossi2021}, with a learned histogram proposal mixed with a
uniform component.  The uniform contribution ensures nonzero support
throughout the vertex domain, while the learned proposal improves efficiency.

\section{Resummation and uncertainty}

At finite volume the shifted expansion has a nonzero convergence radius,
limited by singularities at complex coupling.  The physical point
$\xi=1$ can lie outside this radius.  We use conformal Borel-Leroy
resummation of the available coefficients
\cite{RossiThesis2017,LeGuillouZinnJustin1990},
\begin{align}
\mathcal B_b(t)&=\sum_{k=0}^{K}\frac{a_k t^k}{\Gamma(k+b+1)},\nonumber\\
w(t)&=\frac{\sqrt{1+t/t_0}-1}{\sqrt{1+t/t_0}+1},\nonumber\\
F_K&=\int_0^\infty\!\dd t\,\ee^{-t}t^b
\mathcal B_b^{(K)}(t).
\label{eq:borel}
\end{align}
Here $a_0=f_A^{(0)}$, $a_k=c_k/N$ for $k\geq1$, and
$\mathcal B_b^{(K)}$ is the re-expansion in $w$ through order $K$.
The parameters $b>-1$ and $t_0>0$ are varied to assess the sensitivity
of the finite-order result.  For fixed $(b,t_0)$,
$F_K=\bm\kappa^T\bm a$ is linear in the coefficients.
Each of our $R=128$ independent chains measures all orders, so their
Monte Carlo (MC) errors are correlated.  Chains within each run have
equal lengths.  Deleting one complete chain at a time and recomputing
the shared normalization ratio gives the jackknife covariance
\begin{align}
C&=\frac{R-1}{R}\sum_{r=1}^{R}
(\bm a^{(-r)}-\overline{\bm a}^{(-)})
(\bm a^{(-r)}-\overline{\bm a}^{(-)})^T,\nonumber\\
\sigma_{\rm MC}^2&=\overline{\bm\kappa}^{T}C
\overline{\bm\kappa},
\label{eq:error}
\end{align}
where $\overline{\bm\kappa}$ averages the conformal weights over the
specified window.  Orders zero and one are deterministic.  We report
the sample standard deviation across the window as $\sigma_{\rm scan}$;
it measures parameter sensitivity and is not a calibrated truncation
error.  Parentheses refer to the last displayed digits: $0.12345(67)$ means
$0.12345\pm0.00067$.

\section{Numerical results}

The $3\times3$ cluster permits comparison with ED and deterministic
integration, while the larger lattices share $(U,\mu,\beta)=(3,1,8)$,
away from the particle-hole-symmetric value $\mu=U/2$.  At each size we
vary $\alpha_\sigma$ to test whether distinct expansions approach the
same physical entropy.

\begin{figure*}[t]
\includegraphics[width=\textwidth]{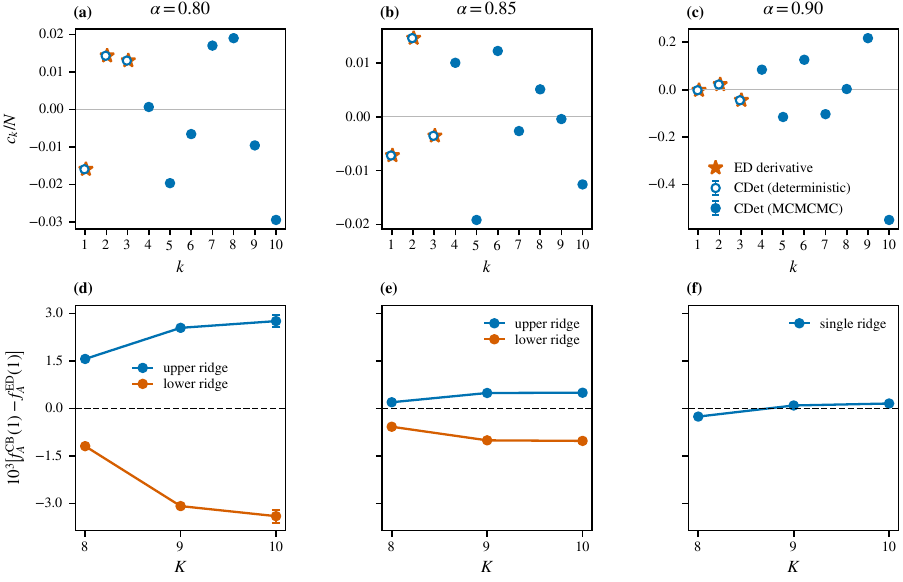}
\caption{\textbf{Small-system benchmark and resummation.}  A $3\times3$
periodic Hubbard lattice with $U=2$, $\mu=0$, $\beta=8$, and $A$ equal to the
first row.  (a)--(c) CDet coefficients through tenth order.  Open circles at
orders 1--3 are deterministic: lattice and replica labels are summed exactly
and the remaining ordered-time integrals use 18-point Gauss--Legendre
quadrature.  Their error bars are the absolute 18-point minus 16-point changes.
Filled circles at orders 4--10 are MCMCMC estimates with chain-level
statistical errors; stars show ED derivatives at orders 1--3.
(d)--(f) Conformal-Borel estimates from the deterministic free and
first-order terms and the MC coefficients at orders 2--10.  Deviations
from ED are shown as $10^3[f_A^{\rm CB}(1)-f_A^{\rm ED}(1)]$ versus
maximum order.  Panels (d) and (e) show the two plateau ridges found
for these shifts; (f) shows the single ridge for $\alpha=0.90$.
Vertical bars at order 10 denote $\sigma_{\rm scan}$.
The dashed line marks the ED endpoint
$\ln Z_2[A]/N=-0.275249813785$.}
\label{fig:three}
\end{figure*}

\emph{Exact-diagonalization benchmark.}
Figure~\ref{fig:three} compares deterministic CDet coefficients with ED
finite-difference derivatives through third order.  For each order $k$, all
lattice and replica labels are summed exactly, leaving a $k$-dimensional
imaginary-time integral.  We partition the time cube into ordered simplices
and apply tensor-product Gauss--Legendre quadrature.  The ED step
$7\times10^{-3}$ lies in the stable finite-difference window; the smallest
tested step is roundoff limited.  With 18 quadrature nodes per transformed
direction, the largest absolute CDet--ED difference over all three shifts and
orders is $8.6\times10^{-9}$.  Increasing the quadrature from 16 to 18
nodes changes the coefficients by at most $1.1\times10^{-7}$.  The panels
also show MCMCMC coefficients from fourth through tenth order, whose
signs and magnitudes depend strongly on the reference shift.

The lower panels of Fig.~\ref{fig:three} compare the resummed values with
ED.  The selected $(b,t_0)$ windows follow regions of local flatness
and order-9-to-order-10 stability.  For $\alpha=0.80$, two distinct
ridges give
$-0.2724919(63)_{\rm MC}$ and $-0.2786529(64)_{\rm MC}$, with scan dispersions
$1.90\times10^{-4}$ and $2.09\times10^{-4}$.  ED lies between these
branches, whose local stability leaves the choice unresolved.
The $\alpha=0.85$ series also has
two local plateaus, $-0.2747537(43)_{\rm MC}$ and
$-0.2762757(44)_{\rm MC}$, separated by much more than either internal
width.  Both shifts retain a substantial branch ambiguity at tenth order.

For $\alpha=0.90$, the selected ridge is continuous.  Its order-10 mean is
\begin{equation}
\begin{aligned}
f_A(1)&=-0.2750946(65)_{\rm MC},\\
\sigma_{\rm scan}&=5.71\times10^{-5}.
\end{aligned}
\label{eq:3result}
\end{equation}
It differs from ED by $1.55\times10^{-4}$, or $0.056\%$.  Although the
order-9 to order-10 drift is only $1.05\times10^{-5}$, the offset
exceeds both the statistical error and the scan dispersion, so
finite-order continuation remains the dominant uncertainty at this
shift.

\begin{figure*}[t]
\includegraphics[width=\textwidth]{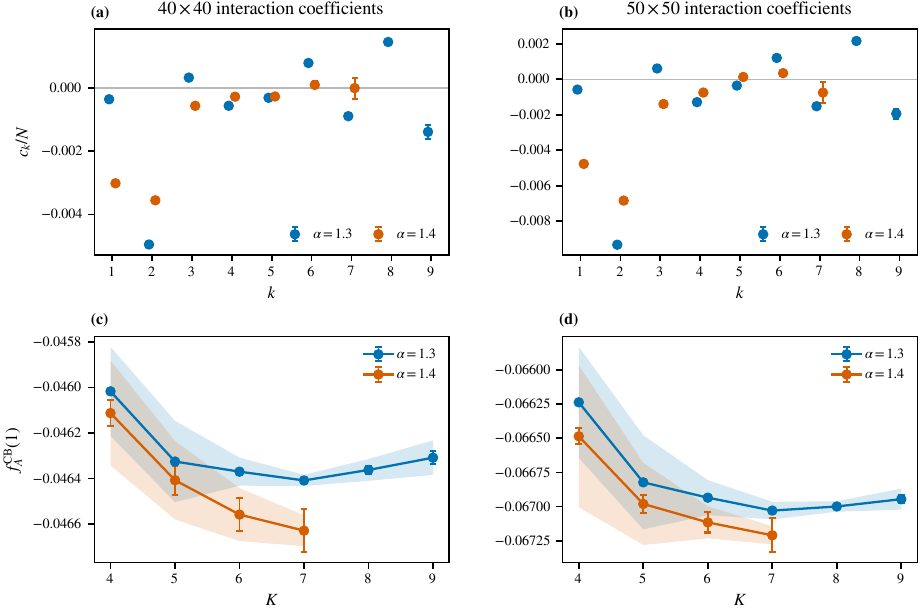}
\caption{\textbf{Large-system shift and truncation stability.}
$U=3$, $\mu=1$, $\beta=8$ on periodic lattices.  The $40\times40$ region is
an origin square of radius $8$ ($|A|=256$); the $50\times50$ region has
radius $13$ ($|A|=676$).  (a),(b) Interaction coefficients for
$\alpha=1.3$ and $1.4$.  (c),(d) Conformal-Borel estimates versus maximum
order, $f_A^{\rm CB}(1)$.  Error bars denote $\sigma_{\rm MC}$ from
Eq.~\eqref{eq:error}; bands denote $\sigma_{\rm scan}$.  The
$\alpha=1.3$ runs use
300 million measured updates per chain.  The $40\times40$,
$\alpha=1.4$ run uses 300 million, and the $50\times50$,
$\alpha=1.4$ run uses 400 million.}
\label{fig:large}
\end{figure*}

\emph{Large finite lattices.}
For both $40\times40$ and $50\times50$, we compare
$\alpha=1.3$ through ninth order with
$\alpha=1.4$ through seventh order.  The common conformal window is
$b=0.5,1,2,3$ and $t_0=5,10,20$ at every truncation shown in
Fig.~\ref{fig:large}.

For $40\times40$, the highest-order results are
\begin{align}
\alpha=1.3,\ K=9:&\quad -0.046308(28)_{\rm MC}(76)_{\rm scan},\nonumber\\
\alpha=1.4,\ K=7:&\quad -0.046629(95)_{\rm MC}(68)_{\rm scan}.
\label{eq:40a13}
\end{align}
Comparing both at the common seventh order gives $-0.046408$ and
$-0.046629$, a difference $2.21\times10^{-4}$.  The $\alpha=1.3$
estimates vary little over orders 7--9, although the scan width
increases again at ninth order.

For $50\times50$, the corresponding values are
\begin{align}
\alpha=1.3,\ K=9:&\quad -0.066945(30)_{\rm MC}(77)_{\rm scan},\nonumber\\
\alpha=1.4,\ K=7:&\quad -0.067210(125)_{\rm MC}(66)_{\rm scan}.
\label{eq:50}
\end{align}
At common order 7 they become $-0.067029$ and $-0.067210$, differing by
$1.81\times10^{-4}$.  The order-7 differences are 2.3 and 1.4 times the
combined MC error for $40\times40$ and $50\times50$, respectively.  The
two expansions thus give nearby endpoint estimates, and their residual
shift dependence provides an empirical scale for the remaining
continuation uncertainty.
The $50\times50$, $\alpha=1.4$ run also stored its Green-function
coefficients with ten fewer mantissa bits, so its comparison with
$\alpha=1.3$ includes this precision change.

\section{Discussion}

Encoding the entangling cut in the free propagator lets the local Hubbard
vertices enter the usual connected-determinant recursions.  The
deterministic comparison with ED checks this construction through third
order.  At higher orders, the small-cluster results leave a measurable
continuation error and, for two shifts, competing branches.  Resolving
these branches and reducing the shift dependence seen on the larger
lattices will require additional coefficients or a better continuation
of the series.

The present calculations reach $50\times50$ by preparing the free
propagators before sampling.  The four replica blocks contain $4N_s^2$
ordered site pairs, where $N_s=N_1N_2$: $10.24$ million for $40\times40$
and $25.0$ million for $50\times50$, a ratio $2.44=(2500/1600)^2$.
At fixed Chebyshev resolution their storage therefore grows as
$O(N_s^2)$, and their preparation involves dense matrix factorizations
and basis transforms with cubic arithmetic cost.  Once the tables are in
memory, a propagator lookup evaluates a fixed-size polynomial, so the
dominant cost per MC step depends on the CDet expansion order.

For comparison, the average sign in reweighted DQMC has the form
$\langle s\rangle_{|w|}=Z/Z_{|w|}=\exp[-\beta N_s\Delta f]$, with
$\Delta f$ the free-energy-density difference between the signed and
absolute-weight ensembles.  A nonzero $\Delta f$ leads to exponentially
increasing effort at fixed relative precision as $\beta N_s$ grows
\cite{TroyerWiese2005}.  Replica estimators can also have poor overlap
between ordinary and sewn ensembles \cite{WangTroyer2014,BroeckerTrebst2014}.
Our expansion samples a finite set of interaction vertices at each
order, without an extensive auxiliary-field sign reweighting factor.
Its statistical cost depends on cancellations in the connected
coefficients; the required order and precision depend on temperature,
coupling, and geometry.  These limitations remain relevant even when
the propagator tables fit in memory.

Two-copy interference and randomized-measurement protocols target the
same moment, $\Tr\rho_A^2$, on a selected spatial region
\cite{Pichler2013,Islam2015,Elben2018}.  Numerical thermometry and entropy
calibration already support Fermi--Hubbard experiments
\cite{Cocchi2017,Xu2025}.  A finite-lattice calculation can match the
experimental size and subsystem geometry, with temperature and filling
specified independently.  Varying the region would give access to
mutual information and the crossover between bulk thermal and boundary
contributions.  Quantitative comparisons of this kind will depend on
reducing the continuation uncertainty identified by the present
benchmarks.

\begin{acknowledgments}
Numerical calculations were performed on Imperial College Research Computing
Service resources.
\end{acknowledgments}

\bibliographystyle{apsrev4-2}
\bibliography{references_expanded_numerics}

\end{document}

%% file: fig_graded_swap_boundary_tikz.tex
\begin{tikzpicture}[
  scale=0.94,
  transform shape,
  >=Latex,
  line cap=round,
  line join=round,
  every node/.style={font=\scriptsize}
]
  \definecolor{sheetlight}{RGB}{244,248,250}
  \definecolor{sheetdeep}{RGB}{218,232,239}
  \definecolor{sheetedge}{RGB}{68,87,101}
  \definecolor{seamblue}{RGB}{0,105,158}
  \definecolor{seamgold}{RGB}{205,83,0}

  \coordinate (uFL) at (0.62,2.35);
  \coordinate (uFR) at (6.55,2.35);
  \coordinate (uBR) at (7.30,2.94);
  \coordinate (uBL) at (1.37,2.94);
  \coordinate (lFL) at (0.62,0.48);
  \coordinate (lFR) at (6.55,0.48);
  \coordinate (lBR) at (7.30,1.07);
  \coordinate (lBL) at (1.37,1.07);

  \fill[black!10,rounded corners=1.4pt]
    ($(uFL)+(0.13,-0.13)$) -- ($(uFR)+(0.13,-0.13)$) --
    ($(uBR)+(0.13,-0.13)$) -- ($(uBL)+(0.13,-0.13)$) -- cycle;
  \fill[black!10,rounded corners=1.4pt]
    ($(lFL)+(0.13,-0.13)$) -- ($(lFR)+(0.13,-0.13)$) --
    ($(lBR)+(0.13,-0.13)$) -- ($(lBL)+(0.13,-0.13)$) -- cycle;
  \fill[sheetdeep!75]
    (uFL) -- (uFR) -- ($(uFR)+(0,-0.09)$) -- ($(uFL)+(0,-0.09)$) -- cycle;
  \fill[sheetdeep!75]
    (lFL) -- (lFR) -- ($(lFR)+(0,-0.09)$) -- ($(lFL)+(0,-0.09)$) -- cycle;

  \shade[top color=sheetlight,bottom color=sheetdeep,
         draw=sheetedge,rounded corners=1.4pt,line width=0.55pt]
    (uFL) -- (uFR) -- (uBR) -- (uBL) -- cycle;
  \shade[top color=sheetlight,bottom color=sheetdeep,
         draw=sheetedge,rounded corners=1.4pt,line width=0.55pt]
    (lFL) -- (lFR) -- (lBR) -- (lBL) -- cycle;

  \foreach \s in {0.28,0.56,0.82} {
    \draw[sheetedge!18,line width=0.35pt]
      ($(uFL)!\s!(uBL)$) -- ($(uFR)!\s!(uBR)$);
    \draw[sheetedge!18,line width=0.35pt]
      ($(lFL)!\s!(lBL)$) -- ($(lFR)!\s!(lBR)$);
  }
  \draw[sheetedge,-{Latex[length=2.1mm]},line width=0.58pt]
    (0.88,2.46) -- (1.28,2.77) node[midway,left=1pt] {$\tau$};
  \draw[sheetedge,-{Latex[length=2.1mm]},line width=0.58pt]
    (0.88,0.59) -- (1.28,0.90) node[midway,left=1pt] {$\tau$};
  \node[anchor=east,font=\footnotesize] at (0.43,2.65) {$\mathcal M_1$};
  \node[anchor=east,font=\footnotesize] at (0.43,0.78) {$\mathcal M_2$};

  \foreach \y in {2.64,0.77} {
    \draw[white,line width=5.0pt] (2.62,\y) -- (5.31,\y);
    \draw[seamblue,line width=1.05pt] (2.62,{\y+0.062}) -- (5.31,{\y+0.062});
    \draw[seamgold,line width=1.05pt] (2.62,{\y-0.062}) -- (5.31,{\y-0.062});
    \fill[sheetedge] (2.62,\y) circle (1.15pt);
    \fill[sheetedge] (5.31,\y) circle (1.15pt);
  }
  \node[font=\footnotesize,fill=sheetlight,inner sep=0.7pt] at (3.97,2.87) {$A$};
  \node[font=\footnotesize,fill=sheetlight,inner sep=0.7pt] at (3.97,1.00) {$A$};
  \node[text=sheetedge] at (2.12,2.40) {$\bar A$};
  \node[text=sheetedge] at (1.92,0.60) {$\bar A$};
  \node[text=sheetedge!80,font=\tiny] at (2.62,3.05) {$\partial A$};
  \node[text=sheetedge!80,font=\tiny] at (5.31,3.05) {$\partial A$};

  \draw[seamblue,-{Latex[length=2.5mm]},line width=1.35pt]
    (5.35,2.66) .. controls (7.73,2.38) and (7.65,1.25) .. (5.35,0.77);

  \draw[seamgold,-{Latex[length=2.5mm]},line width=1.35pt]
    (2.59,0.77) .. controls (-0.08,1.02) and (-0.05,2.23) .. (2.59,2.65);

  \draw[sheetedge!50,densely dashed,line width=0.45pt]
    (2.62,2.64) -- (2.62,0.77);
  \draw[sheetedge!50,densely dashed,line width=0.45pt]
    (5.31,2.64) -- (5.31,0.77);
  \node[font=\footnotesize,text=sheetedge] at (3.97,1.78)
    {$S_A^{\mathrm{gr}}$};
  \node[font=\tiny,text=seamgold,anchor=east] at (3.78,1.51)
    {$2\longrightarrow-1$};
  \node[font=\tiny,text=seamblue,anchor=west] at (4.16,1.51)
    {$1\longrightarrow2$};
  \node[font=\tiny,text=sheetedge!75] at (3.97,1.30)
    {cross-sheet sewing};

  \node[anchor=north,align=center,inner sep=0pt] at (3.97,0.24) {%
    $\begin{aligned}
      i\in A:\quad
      &c_A^{(1)}(\beta^-)=c_A^{(2)}(0^+),\qquad
       c_A^{(2)}(\beta^-)=-c_A^{(1)}(0^+),\\[4.5pt]
      i\in\bar A:\quad
      &c_{\bar A}^{(r)}(\beta^-)=-c_{\bar A}^{(r)}(0^+).
    \end{aligned}$};
\end{tikzpicture}